\documentclass[11p,reqno]{amsart}
\usepackage[T1]{fontenc}
\usepackage{graphicx}
\usepackage{amssymb,amsthm,amsmath,mathrsfs,bm,braket,marginnote}
\usepackage{enumerate}
\usepackage{appendix}
\usepackage{bbm}
\usepackage{eufrak}
\usepackage{mathrsfs}
\usepackage[colorlinks=true, pdfstartview=FitV, linkcolor=blue, citecolor=blue, urlcolor=blue]{hyperref}
\usepackage{pgf}
\usepackage{pgfplots}
\usepackage{tikz}
\usetikzlibrary{arrows,calc}
\usepackage{verbatim}
\usetikzlibrary{decorations.pathreplacing,decorations.pathmorphing}

\numberwithin{equation}{section}
\newcommand{\pf}{\mbox{pf}}
\newcommand{\Pf}{\mbox{Pf}}

 \reversemarginpar
\begin{document}
\title[Matrix integral solutions to the related Leznov lattice equations]{Matrix integral solutions to the related Leznov lattice equations}

\author{Bo-Jian Shen}
\address{School of Mathematical Sciences, Shanghai Jiao Tong University, People's Republic of China.}
\email{JOHN-EINSTEIN@sjtu.edu.cn}

\author{Guo-Fu Yu}
\address{School of Mathematical Sciences, Shanghai Jiao Tong University, People's Republic of China.}
\email{gfyu@sjtu.edu.cn}

\date{}
\dedicatory{}

\begin{abstract}
Matrix integrals used in random matrix theory for the study of eigenvalues of matrix ensembles have been shown to provide $ \tau $-functions for several hierarchies of integrable equations. In this paper, we construct the matrix integral solutions to the Leznov lattice equation, semi-discrete and full-discrete version and the Pfaffianized Leznov lattice systems, respectively. We demonstrate that the partition function of Jacobi unitary ensemble is a solution
to the semi-discrete Leznov lattice and the partition function of Jacobi orthogonal/symplectic ensemble gives solutions of the Pfaffianized Leznov lattice.
\end{abstract}
\maketitle

\section{Introduction}
\setcounter{equation}{0}
The studies in the connections between matrix models and integrable systems originated from the context of string theory and AdS/CFT correspondence \cite{gerasimov90}. The matrix models were used to evaluate non-perturbative correlators in string models constructed from conformal field theories, while at the same time, they were used to provide different tau functions for the integrable equations and hierarchies. During the courses of these studies, the famous integrable  KP and 1d-Toda chain were firstly connected with the Hermitian matrix model with unitary invariance \cite{makeenko91,adler97}. Later on, with the development of random matrix theory, many different matrix models were considered during the late 1990s. The most general case were considered in \cite{adler99} about two-matrix models and it was shown that the time-dependent partition function of coupled two matrix models is the tau function of general 2d-Toda hierarchy \cite{adler99}. According to the Kyoto school's classification about integrable hierarchies \cite{jimbo83}, the equations and hierarchies mentioned above belong to $A_\infty$ type, and therefore, it is natural to consider the other kinds of integrable hierarchies afterwards. The matrix integral solution of BKP hierarchy was shown to be the time-dependent partition function of Bures ensemble \cite{hu17,orlov16} and the one of DKP (or Pfaff lattice) hierarchy were demonstrated to be the partition function of orthogonal ensemble or symplectic ensemble \cite{orlov16,adler02,vandeleur01,kakei00}. Very recently, the matrix integral solution of CKP hierarchy were shown to be the time-dependent partition function of Cauchy two-matrix model \cite{CS}, and thus complete a list of connections between typical integrable hierarchies (we mean the AKP, BKP, CKP and DKP equation in the Kyoto school's classification) and the matrix models.

The discrete integrable systems also play important roles in integrable theory and their connections with matrix models are interesting to be discussed. For example, in \cite{HZL}, the matrix integral solutions of several integrable differential-difference systems were considered, whose tau functions were shown to be related to the integrals of the eigenvalues in the form
\begin{equation}
\int_{\gamma^n} \prod_{1 \leq j<k\leq N}|x_j-x_k|^\beta \exp\left(\sum_{j=1}^Nf(x_j)\right)dx_1\dots dx_N,\label{1.1}
\end{equation}
where the Dyson index $\beta=1,\,2,\, 4$ correspond to the orthogonal, unitary, and symplectic ensembles respectively and $\exp(f(x))$ is a weight function with respect to the integral contour $\gamma$. Moreover, in \cite{lafortune16}, the matrix integral solutions to the discrete KP hierarchy and its Pfaffianized version were considered, which provides us an idea to extend the original equations to Pfaffianized systems, to connect the tau function of the original one with $\beta=2$ with the tau functions of the Pfaffianized one with $\beta=1,\,4$.

In this article, we'd like to extend the relation between integrals of the form (\ref{1.1}) and integrable systems by obtaining matrix integral solutions to the semi-discrete and full-discrete Leznov lattice.
The nonlinear two-dimensional (2D) Leznov lattice  \cite{lez2000}
\begin{subequations}
\begin{align}
& \frac{\partial p(n)}{\partial y}=\theta(n+1)-\theta(n-1),\label{lez-1}\\
& \frac{\partial^2}{\partial x\partial y}\ln\theta(n)=\theta(n+1)p(n+1)-2\theta(n)p(n)+\theta(n-1)p(n-1), \label{lez-2}
\end{align}
\end{subequations}
which is a special case of the so-called UToda $(m_1,m_2)$ system with
$m_1=1,m_2=2$. If we set the variable transformations $\,
a(n)=p(n+1),c(n)=\theta(n+1)$, then (\ref{lez-1}) and (\ref{lez-2}) can be
transformed into
\begin{subequations}
\begin{align}
&a_y(n)=c(n+1)-c(n-1),\\
&b_y(n)=a(n-1)c(n-1)-a(n)c(n),\\
&c_x(n)=c(n)[b(n)-b(n+1)],
\end{align}
\end{subequations}
which is a two-dimensional generalization of the Blaszak-Marciniak
lattice \cite{BM94,HuZhu98}.
By the dependent variable transformation \cite{HTam}
\begin{equation*}
\theta(n)=\frac{f(n+1)f(n-1)}{f(n)^2},\quad p(n)=\frac{1}{2}\frac{D_xD_yf(n)\cdot f(n)}{f(n+1)f(n-1)},
\end{equation*}
the Leznov lattice (\ref{lez-1})-(\ref{lez-2}) can be transformed into a quadric linear form
\begin{eqnarray}
\frac{1}{2}D_y(D_xD_yf(n)\cdot f(n))\cdot (e^{D_n}f(n)\cdot
f(n))=2\sinh(D_n)(e^{D_n}f(n)\cdot f(n))\cdot f^2(n).\label{quadr}
\end{eqnarray}
To decouple (\ref{quadr}) into  the bilinear form, we need to introduce an auxiliary variable $z$ to obtain
\begin{subequations}
\begin{align}
& (D_yD_z-2\,\mathrm{e}^{D_n}+2)f(n)\cdot f(n)=0,\label{le1}\\
& (D_yD_x-2D_z \mathrm{e}^{D_n})f(n)\cdot f(n)=0,\label{le2}
\end{align}
\end{subequations}
where the Hirota's bilinear differential operator $D_y^mD_t^k$ and the bilinear difference operator $\exp(\delta D_n)$ are defined \cite{H2}, respectively, by
\begin{eqnarray*}
&& D_y^m D_t^k a\cdot b \equiv (\frac{\partial}{\partial y}- \frac{\partial}{\partial {y'}})^m  (\frac{\partial}{\partial t}- \frac{\partial}{\partial {t'}})^k  a(y,t)b(y',t')|_{y'=y,t'=t},
\\ &&\exp(\delta D_n)a(n)\cdot b(n) \equiv \exp\Big[\delta \Big(\frac{\partial}{\partial n}-\frac{\partial}{\partial {n'}}\Big)\Big]  a(n)b(n')|_{n'=n}  = a(n+\delta)b(n-\delta).
\end{eqnarray*}
Moreover, by making use of the direct integrable discretization method proposed by Hirota \cite{H1}, the integrable difference Leznov lattice equations were constructed in \cite{JDEA09} and the discrete integrability were demonstrated by the B\"acklund transformation and Lax pair. The advantage of this method is to find the solutions easily. It is known that the original Leznov lattice has Wronskian determinant solutions, and therefore the solutions of integrable semi(full)-discrete Leznov lattices inherit the determinant structure, which means both of them have Casorati determinant solutions \cite{GHX}. The Pfaffianization procedure \cite{ohta,Gilson1,Gilson2,Gilson3} was also applied to the Leznov lattice and demonstrated that Pfaffianized Leznov lattice equation has solutions which could be expressed by Pfaffian \cite{GHX}.

Motivated by the previous study of Leznov lattice and the importance of the matrix integrals, in this paper, we'd like to consider the matrix integral solutions to the Leznov lattice and its Pfaffianized version. The rest of this paper is organized as follows. In section 2, we present the Andr\'eief formula  and  matrix integral formulae that solve the Leznov lattice and its discrete versions. We show that the partition function of Jacobi unitary ensemble (JUE) in (\ref{jue}) is a solution to the semi-discrete Leznov lattice in the $y$-direction. While in section 3, we demonstrate that the partition function of Jacobi orthogonal/symplectic ensemble  form solutions to the  Pfaffianized semi-discrete Leznov lattice  in $y$-direction and the Pfaffianized Leznov lattice system. Finally, conclusion and discussion are given in section 4.

\section{Matrix integral solutions to the Leznov lattice and discrete Leznov lattices}
\subsection{Matrix integral solution to the Leznov lattice}
In \cite{zhao}, it was shown that the bilinear Leznov lattice equation \eqref{le1}-\eqref{le2} has the Casorati determinant solution
\begin{align}\label{Cast}
f(n)
=\left|\begin{array}{cccccccc}\phi_1(n) & \phi_1(n+1)& \cdots &\phi_1(n+N-1)\\
\phi_2(n) & \phi_2(n+1) & \cdots & \phi_2(n+N-1)\\
\vdots & \vdots  &        & \vdots
\\
\phi_N(n) & \phi_N(n+1)  & \cdots & \phi_N(n+N-1)
\end{array}
\right|,
\end{align}
where $\{\phi_i(n):=\phi_i(n,x,y,z),\,i=1,2,\cdots, N\}$ satisfy the dispersion relations
\begin{align}
	\frac{\partial \phi_j(n)}{\partial y}=\phi_j(n+1),\quad
\frac{\partial \phi_j(n)}{\partial z}=-\phi_j(n-1),\quad
\frac{\partial \phi_j(n)}{\partial x}=-\phi_j(n-2).
\end{align}
It should be emphasized that if we consider the Casorati determinant of the form \eqref{Cast} with the seed function $\phi_i(n,x,y)$, which is only dependent on variables $n,\,x,\,y$, then the determinant is also the tau function of Leznov lattice in quadric form \eqref{quadr}, reflecting the fact that $z$ is only an auxiliary variable and wouldn't change the properties of solutions. Therefore, in the latter use, we would like to say the tau functions with variable $z$ are the solutions of the decoupled bilinear forms and the ones without $z$ are the solutions of the quadric forms or nonlinear forms.

A special solution solving the above relations is given as
\begin{align*}
\phi_i(n)=\int_\gamma t^{n+i-1}e^{\eta(x,y,z,t)}dt,
\end{align*}
with $\eta(x,y,z,t)=yt-zt^{-1}-xt^{-2}+\eta_0(t)$ and $\eta_0(t)$ is a weight function with respect to the integral contour $\gamma$. Usually, in integrable theory, the variable $y$ is regarded as the $t_1$-time flow and $x$ is regarded as the $t_{-2}$-time flow.  Considering $z$ is an auxiliary variable, the Casorati determinant solution \eqref{Cast} with $z=0$ solves multi-linear Leznov lattice equation \eqref{quadr}.

By making use of the Andr\'eief formula
\begin{align}
\frac{1}{N!}\int_{\gamma^n} \det[\phi_i(x_j)]_{i,j=1}^{N} \det[\psi_i(x_j)]_{i,j=1}^N\prod_{i=1}^N \omega(x_i)dx_i=\det\left[\int_\gamma \phi_i(x)\psi_j(x)\omega(x)dx\right]_{i,j=1}^N,\label{dB.1}
\end{align}
one can write the Casorati determinant as a matrix integral form and obtain
\begin{align}
\det\limits_{1\leq i,j\leq N}\Big[\int_\gamma t^{n+i+j-2}e^{\eta(x,y,t)}dt\Big]=\frac{1}{N!}\int_{\gamma^n}\prod_{1\leq i<j\leq N}|t_i-t_j|^2\prod_{i=1}^Nt_i^n \exp[\eta(x,y,t_i)]dt_i,\notag
\end{align}
where $\eta(x,y,t)=yt-xt^{-2}+\eta_0(t)$. Thus we conclude that the partition function of the Hermitian matrix model with unitary invariance gives the tau function of  Leznov lattice equation with dispersion relations given in the weight.

\subsection{Matrix integral solution to the semi-discrete Leznov lattice in $y$-direction}\hspace*{\fill}\\
The semi-discrete version of the Leznov lattice in the $y$-direction is given as \cite{GHX}
\begin{subequations}
\begin{align}
& D_ze^{D_m}f(n,m)\cdot f(n,m)=(2e^{D_m+D_n}-2e^{D_m})f(n,m)\cdot f(n,m),\label{ey.1}\\
& (D_xe^{D_m}-2D_ze^{D_m+D_n})f(n,m)\cdot f(n,m)=0,\label{ey.2}
\end{align}
\end{subequations}\label{ey}
where we denote $f(n,m)=f(n,x,m,z)$ for simplicity and $m$ takes the place of $y$ as a discrete variable.
By using the bilinear identity
\begin{align*}
\sinh (D_m)(D_ze^{D_m+D_n}a\cdot a)\cdot(e^{D_m+D_n}a\cdot
a)=\sinh(D_n+D_m)(D_ze^{D_m}a\cdot a)\cdot (e^{D_m}a\cdot
a),
\end{align*}
we can derive the multi-linear equation
\begin{align}
\frac 12\sinh (D_m)[D_xe^{D_m}f\cdot f]\cdot [e^{D_n+D_m}f\cdot
f]=2\sinh (D_n+D_m)[e^{D_m+D_n}f\cdot f]\cdot [e^{D_m}f\cdot
f]
\end{align}
from bilinear equations \eqref{ey.1}-\eqref{ey.2}.
Moreover, this quadric linear equation could be regarded as the discrete version of the multi-linear equation \eqref{quadr}.
In \cite{GHX}, the following Casorati determinant solution was obtained
\begin{align}
f(n,m)=\left|\begin{array}{cccc}
\phi_1(n,m)&\phi_1(n+1,m)&\dots& \phi_1(n+N-1,m)\\
\phi_2(n,m)&\phi_2(n+1,m)&\dots& \phi_2(n+N-1,m)\\
\vdots& \vdots& \ddots& \vdots \\
\phi_N(n,m)&\phi_N(n+1,m)&\dots& \phi_N(n+N-1,m)\\
\end{array}
\right|,\label{ca.y}
\end{align}
where $\{\phi_{i}(k,m):=\phi_i(k,x,m,z),\,k=n,\,n+1,\,\dots,\,n+N-1\}$ satisfy the dispersion relations
\begin{subequations}
\begin{align}
\partial_x\phi_i(k,m)&= -4\phi_i(k-2,m)+4\phi_i(k-1,m),\label{dry.1}\\
\partial_z\phi_i(k,m)&=-2\phi_i(k-1,m), \label{dry.2}\\
\phi_i(k,m)&=\phi_i(k,m+2)-\phi_i(k+1,m+2). \label{dry.3}
\end{align}
\end{subequations}
To seek for the solitons, one can consider the seed functions
\begin{align*}
\phi_i(n,x,m,z)&=c_ip_i^n(1-p_i)^{-\frac{m}{2}}\exp\left(-2p_i^{-1}z+(-4p_i^{-2}+4p_i^{-1})x\right)\notag\\
&+d_iq_i^n(1-q_i)^{-\frac{m}{2}}\exp({-2q_i^{-1}z+(-4q_i^{-2}+4q_i^{-1})x})
\end{align*}
motivated by the dispersion relations \eqref{dry.1}-\eqref{dry.3} and $p_i,\,q_i,\,c_i,\,d_i$ can be taken as arbitrary constants, which usually correspond to the wave numbers and phase parameters of the $i-$th soliton, respectively.
In what follows, we'd like to show the partition function of Jacobi unitary ensemble (JUE) can act as the tau function of the semi-discrete Leznov lattice in the $y$-direction. The approach is to obtain a particular case of the Casorati determinant solution (\ref{ca.y}), and show it is equivalent to the partition function of JUE. It is easy to check that the $ \phi_i(n,m) $ defined by
\begin{align}
\phi_i(n,x,m,z)=\int_0^{1}(1-t)^{-\frac{m}{2}}t^{n+i-1}e^{\eta_1(x,z,t)}dt
\end{align}
satisfy the dispersion relation (\ref{dry.1})-(\ref{dry.3}). Here $\eta_1(x,z,t)=-2zt^{-1}+(-4t^{-2}+4t^{-1})x+\eta_0(t)$ with
$\eta_0(t)$ taken to ensure the convergence of the integral. Noting that $z$ is an auxiliary variable and it is not essential in the quadric form, we ignore this parameter in the following formula.
Again, by making use of Andrei\'ef formula \eqref{dB.1}, one can see
\begin{align}
\det\limits_{1\leq i,j\leq N}\left[\int_0^1(1-t)^{-\frac{m}{2}}t^{n+i+j-2}e^{\eta_1(x,t)}dt\right]
=\frac{1}{N!}\int_{[0,1]^N}\prod_{1\leq i<j\leq N}|t_i-t_j|^2\prod_{i=1}^N (1-t_i)^{-\frac{m}{2}}t_i^n e^{\eta_1(x,t_i)}dt_i,\label{jue}
\end{align}
which is the partition function of JUE with correspondence to the weight function $(1-t)^{-m/2}t^n$.

\subsection{Matrix integral solutions to the semi-discrete Leznov lattice in $x$-direction}\hspace*{\fill}\\
The bilinear semi-discrete version of the Leznov lattice in the $x$-direction is
\begin{subequations}\label{ex}
\begin{align}
& D_yD_zf\cdot f=(2e^{D_n}-2)f\cdot f,\label{ex.1}\\
& (D_ye^{D_k}-2D_ze^{D_k+D_n})f\cdot f=0.\label{ex.2}
\end{align}
\end{subequations}
By use of the bilinear identity
\begin{equation*}
\sinh (D_n+D_k)(D_yD_za\cdot a)\cdot a^2=D_y(D_ze^{D_n+D_k}a\cdot
a)\cdot (e^{D_n+D_k}a\cdot a),
\end{equation*}
we get the multi-linear equation
\begin{equation}\label{dxx}
\frac 12D_y[D_ye^{D_k}f\cdot f]\cdot [e^{D_n+D_k}f\cdot f]=2\sinh
(D_n+D_k)[e^{D_n}f\cdot f]\cdot f^2,
\end{equation}
that is an $x-$direction discrete version of \eqref{quadr}.
The solution of the semi-discrete version of the Leznov lattice (\ref{ex.1})-(\ref{ex.2}) can be written as the Casorati determinant \cite{GHX}
\begin{align}\label{sol1}
f(n,l)=\left|
\begin{array}{cccc}
\psi_1(n,l)&\psi_1(n+1,l)&\dots& \psi_1(n+N-1,l)\\
\psi_2(n,l)&\psi_2(n+1,l)&\dots& \psi_2(n+N-1,l)\\
\vdots& \vdots& \ddots& \vdots \\
\psi_N(n,l)&\psi_N(n+1,l)&\dots& \psi_N(n+N-1,l)\\
\end{array}
\right|,
\end{align}
where $\{\psi(k,l):=\psi_i(k,l,y,z),\,k=n,\,n+1,\,\dots,\,n+N-1\} $ satisfy the dispersion relations
\begin{subequations}
\begin{align}
\partial_y\psi_i(k,l)&=-\psi_i(k-1,l),\label{drx.1}\\
\partial_z\psi_i(k,l)&=\psi_i(k+1,l),\label{drx.2}\\
\psi_i(k,l)&=-2\psi_i(k+2,l+2)+\psi_i(k,l+2).\label{drx.3}
\end{align}
\end{subequations}
The substitution the Casorati determinants into bilinear equations \eqref{ex.1}-\eqref{ex.2} leads to the Pl\"ucker identity and therefore \eqref{sol1} is a solution of semi-discrete Leznov lattice in $x$-direction. A special choice of the seed function
\begin{equation*}
\psi_i(n,l,y)=\int_{0}^{\sqrt{2}/2}(1-2t^2)^{-\frac{l}{2}}t^{n+i-1}e^{\eta_2(t,y)}dt,\quad \eta_2(t,y)=-yt^{-1}+\eta_0(t),
\end{equation*}
solves dispersion relations \eqref{drx.1}-\eqref{drx.3} with ignorance of the auxiliary variable $z$. In the case that $1-2t^2>0$, we can put this term into the exponential term and express the seed functions as the form
\begin{align*}
\psi_i(n,l,y)=\int_0^{\sqrt{2}/2}t^{n+i-1}e^{\tilde{\eta}_2(t,l,y)}dt,\quad \tilde{\eta}_2(t,l,y)=-\frac{l}{2}\log(1-2t^2)-yt^{-1}+\eta_0(t).
\end{align*}
The Andrei\'ef formula \eqref{dB.1} is used to show a matrix integral solution to the semi-discrete Leznov lattice in $x$-direction
\begin{align*}
f(n,l,y)=\det\limits_{1\leq i,j\leq N}\left[
\int_{0}^{\sqrt{2}/2} t^{n+i+j-2}e^{\tilde{\eta}_2(t,l,y)}dt
\right]=\frac{1}{N!}\int_{[0,\sqrt{2}/2]^N}\prod_{1\leq i<j\leq N}|t_i-t_j|^2\prod_{i=1}^N t_i^{n}e^{\tilde{\eta}_2(t_i,l,y)}dt_i,
\end{align*}
which solves semi-discrete equation \eqref{dxx}.

\subsection{Matrix integral solution to the full-discrete Leznov lattice}\hspace*{\fill}\\
The full-discrete version of the Leznov lattice is
\begin{subequations}\label{efqn}
\begin{align}
& D_ze^{D_{m}}f\cdot f=(2e^{D_m+D_n}-2e^{D_m})f\cdot f,\label{ef.1}\\
 &[\sinh(D_m)\sinh(D_k)-2D_ze^{D_n+D_m+D_k}]f\cdot f=0.\label{ef.2}
\end{align}
\end{subequations}
Based on the bilinear operator identity
\begin{equation*}
\sinh (D_m+D_n+D_k)(D_ze^{D_m}a\cdot a)\cdot(e^{D_m}a\cdot
a)=\sinh(D_m)(D_ze^{D_m+D_n+D_k}a\cdot a)\cdot
(e^{D_m+D_n+D_k}a\cdot a),
\end{equation*}
we approach the full-discrete multi-linear equation
\begin{equation*}
\frac 12\sinh (D_m)[\sinh (D_m)\sinh (D_k)f\cdot f]\cdot
[e^{D_m+D_n+D_k}f\cdot f]=2\sinh (D_m+D_n+D_k)(e^{D_m+D_n}f\cdot
f)\cdot (e^{D_m}f\cdot f).
\end{equation*}
Thus we view the continuous variable $z$ in \eqref{efqn} as an auxiliary variable. The Casorati determinant solution of the full-discrete Leznov lattice \eqref{efqn} is expressed by
\begin{align}
f_{m,n,k}=\left|\begin{array}{cccc}
\phi_1(m,k,n,z)&\phi_1(m,k,n+1,z)&\dots& \phi_1(m,k,n+N-1,z)\\
\phi_2(m,k,n,z)&\phi_2(m,k,n+1,z)&\dots& \phi_2(m,k,n+N-1,z)\\
\vdots& \vdots& \ddots& \vdots \\
\phi_N(m,k,n,z)&\phi_N(m,k,n+1,z)&\dots& \phi_N(m,k,n+N-1,z)\\
\end{array}
\right|,\label{ca.f}
\end{align}
where $\{\phi_i(m,k,n,z),\, i=n,\,n+1,\,\dots,\,n+N-1\}$ satisfy the following dispersion relations:
\begin{subequations}
\begin{align}
\phi_i(m-2,k,n,z)&=\phi_i(m,k,n,z)+\phi_i(m,k,n+1,z),\label{drf.1}\\
\partial_z\phi_i(m,k,n,z)&=2\phi_i(m,k,n-1,z),\label{drf.2}\\
\phi_i(m,k+2,n,z)&=\phi_i(m,k,n,z)-8\phi_i(m,k,n-1,z)-8\phi_i(m,k,n-2,z).\label{drf.3}
\end{align}
\end{subequations}
We take a special choice of the seed function as
\begin{align*}
&\phi_i(m,n,k)=\int_{-1}^{0} (1+t)^{-\frac{m}{2}}t^{n+i-1}\Big(1-\frac{8}{t} -\frac{8}{t^2}\Big)^{\frac{k}{2}}\,\exp(\eta_3 (t,z))dt,\notag\\
&\eta_3 (t,z)=\frac{2z}{t}+\eta_0(t).
\end{align*}
Here $\eta_0(t)$ is an arbitrary function to ensure the convergence of the integral. It is easy to verify that $ \phi_i(m,n,k) $ above satisfies the dispersion relations (\ref{drf.1})-(\ref{drf.3}).
Then, using the identity (\ref{dB.1}) and factoring the integrand as the multiplication of two determinants, we get
\begin{align}
f_{m,n,k}&=\det\left[\int_{-1}^{0} (1+t)^{-\frac{m}{2}}t^{n+i+j-2}(1-\frac{8}{t} -\frac{8}{t^2})^{\frac{k}{2}}\exp{(\eta_3 (t))}dt\right]_{i,j=1\dots N}\notag\\
&=\frac{1}{N!} \int_{[-1,0]^N}\det[t_j^{i-1}]_{i,j=1}^N\cdot \det\Big[t_j^{n+i-1}(1+t_j)^{-\frac{m}{2}}
\Big(1-\frac{8}{t_j}-\frac{8}{t_j^2}\Big)^{\frac{k}{2}}\exp[\eta_3(t_j)]\Big]_{i,j=1}^Ndt_j\notag\\
&=\frac{1}{N!} \int_{[-1,0]^N} \prod_{i=1}^N t_i^n(1+t_i)^{-\frac{m}{2}}\left(1-\frac{8}{t_i}-\frac{8}{t_i^2}\right)^{\frac{k}{2}}  \prod_{1\le i<j\le N} |t_i-t_j|^{2}\prod_{i=1}^N\exp\left[\eta_3(t_i,z)\right]dt_i.
\end{align}
Thus we obtain the matrix integral solution to the full-discrete Leznov lattice.

\section{Matrix integral solutions to the Pfaffianized version of the semi-discrete Leznov lattice in the $y$-direction}
In this section, we will show that the partition functions of the orthogonal and symplectic ensemble can be regarded as the tau  solution of the Pfaffianized version of the semi-discrete Leznov lattice in the $y$-direction (\ref{ey.1})-(\ref{ey.2}). We start from recalling some facts about Pfaffians.

As known in \cite[\S 2]{H2}, a Pfaffian $\pf(1,2,\dots ,2N)$ is defined recursively by
\begin{equation*}
\pf(1,2,\dots,2N)= \sum_{j=2}^{2N}(-1)^{j}\pf(1,j)\pf(2,3,\dots,\hat{j},\dots,2N),
\end{equation*}
where $\pf(i,j) = -\pf(j,i)$ and $ \hat{j} $ means that the index $j$ is omitted. For any given $ 2N\times 2N $ antisymmetric matrix $ A_{2N} = (a_{ij})_{1\leq i,j\leq 2N} $, the Pfaffian associated with $ A_{2N} $ is defined as
\begin{eqnarray*}
\Pf[A_{2N}]= \pf(1,2,\dots,2N),
\end{eqnarray*}
with $\pf(i,j) = -\pf(j,i) = a_{ij}$. In fact, Pfaffian is closely related to determinant. A determinant of $ n- $th degree $ \det|b(j,k)|_{1\leq j,k\leq n} $ can be expressed by means of a Pfaffian of $ 2n- $th degree $\pf(1,2,\dots,n,n^*,\dots,2^*,1^*)$ as \cite{H2}
\begin{equation*}
\det|b(j,k)|_{1\leq j,k\leq n}=\pf(1,2,\dots,n,n^*,\dots,2^*,1^*)
\end{equation*}
with entries defined by
\begin{equation*}
\pf(j,k)= \pf(j^*,k^*)=0,\quad \pf(j,k^*)=b(j,k).
\end{equation*}

\subsection{matrix integral solutions of the Pfaffianized semi-discrete Leznov lattice}
The Pfaffianized version of the semi-discrete Leznov lattice in the $y$-direction was obtained in \cite{GHX} by using the procedure of Pfaffianization proposed by Hirota and Ohta \cite{RH}. It takes the form of the following four coupled equations
\begin{subequations}
\begin{align}
&D_zf^n_m\cdot f^n_{m-2}+2f^n_mf^n_{m-2}-2f^{n+1}_mf^{n-1}_{m-2}+g^n_{m-2}h_m^n=0,\label{pfLS-1}\\
&D_xf^n_m\cdot f^n_{m-2}-2f^{n+1}_{m,z}f^{n-1}_{m-2}+2f^{n+1}_mf^{n-1}_{m-2,z}=-2D_zg^n_{m-2}h^n_m+8g^n_{m-2}h^n_m,\\
&D_zf^n_m\cdot g^{n-1}_{m-2}+2f^n_mg^{n-1}_{m=2}+2f^{n-1}_mg^n_{m-2}-2f^{n-1}_{m-2}f^n_m=0,\\
&D_zh^n_m\cdot f^{n-1}_{m-2}+2h^n_mf^{n-1}_{m-2}+2h^{n-1}_mf^n_{m-2}-2h^{n-1}_mf^n_m=0. \label{pfLS-4}
\end{align}
\end{subequations}
It is remarkable the system above has the Pfaffian solutions
\begin{eqnarray*}
&&f^n_m=\pf(1,2,\dots,N)^n_m,\\
&&g_m^n=\pf(2,\dots,N-1)^n_m,\\
&&h_m^n=\pf(0,1,\dots,N,N+1)^n_m
\end{eqnarray*}
supposed that $N$ is an even integer larger than $0$ and the entries of the Pfaffians are chosen to satisfy relations
\begin{subequations}
\begin{align}
&\pf(i,j)^n_m-\pf(i,j)^n_{m-2}=\pf(i+1,j)^n_m+\pf(i,j+1)^n_m-\pf(i+1,j+1)^n_m,\label{drpf1}\\
&\frac{\partial}{\partial x}\pf(i,j)^n_m=-4\pf(i-2,j)^n_m-4\pf(i,j-2)^n_m+4\pf(i,j-1)^n_m+4\pf(i-1,j)^n_m,\\
&\frac{\partial}{\partial z}\pf(i,j)^n_m=-2\pf(i-1,j)^n_m-2\pf(i,j-1)^n_m.\label{drpf3}
\end{align}
\end{subequations}
In the following, we'd like to present two different kinds matrix integral solutions to the coupled Leznov lattice \eqref{pfLS-1}-\eqref{pfLS-4}.

\subsubsection{ Matrix integral solutions (I)}

First, we consider the matrix integral related to the orthogonal ensemble with even size $N$, which is of the form
\begin{align*}
&\mathbb{Z}_{N}^{(\beta=1)}=\frac{1}{N!}\int_{\gamma^{N}}
\prod_{1\le i<j\le N}|t_i-t_j|\prod_{i=1}^{N}\omega(t_i)dt_i.
\end{align*}
It is well know the above equation can be written as a Pfaffian in the virtue of the de Bruijn integral formula \cite{dB}
\begin{align*}
&\frac{1}{N!}\int_{\gamma^n} \left|\det\limits_{1\leq i,j\leq N}[\phi_i(t_j)]\right|\prod_{i=1}^N\omega(t_i)dt_i=\mathrm{Pf}\left[\int_{\gamma^2}\text{sgn}(s-t)(\phi_i(t)\phi_j(s)-\phi_i(s)\phi_j(t))\omega(s)\omega(t)dsdt\right]_{i,j=1}^{N}
\end{align*}
and therefore one can consider the Pfaffian elements
\begin{eqnarray*}
\pf(i,j)=\int_0^{1} \int_s^{1} (s^{i-1}t^{j-1}-s^{j-1}t^{i-1})\omega(s)\omega(t)dtds.
\end{eqnarray*}
By considering that the weight function is parameter dependent, i.e. $\omega(s):=\omega(s;x,z,m,n)$ such that
\begin{align*}
\omega(s;x,z,m,n)=s^n(1-s)^{-m/2}\exp(\eta_1(x,z,s)),\quad \eta_1(x,z,s)=-2zs^{-1}+(-4s^{-2}+4s^{-1})x+\eta_0(s)
\end{align*}
where $\eta_0(s)$ is taken to ensure the convergence.
Obviously, the Pfaffian entry defined above satisfies the dispersion relation \eqref{drpf1}-(\ref{drpf3}). Therefore, by using the de Bruijn formula backwards, we can check that
\begin{align*}
& f^n_m=\pf(1,2,\dots,N)^n_m=\mathbb{Z}_N^{(\beta=1)}(x,z,m,n),\\
&g^n_m=\pf(2,\dots,N-1)^n_m=\mathbb{Z}_{N-2}^{(\beta=1)}(x,z,m,n+1),\\
& h^n_m=\pf(0,1,\dots,N,N+1)^n_m=\mathbb{Z}_{N+2}^{(\beta=1)}(x,z,m,n-1).
\end{align*}
Noting that the partition function expression of $f_m^n$ could be written as
\begin{align*}
\mathbb{Z}_N^{(\beta=1)}(x,z,m,n)=\frac{1}{N!}\int_{\gamma^N}\prod_{1\leq i<j\leq N}|t_i-t_j|\prod_{i=1}^N t_i^n (1-t_i)^{-m/2}\exp(\eta_1(x,z,t_i))dt_i,
\end{align*}
and this is the parameter dependent partition function of Jacobi Orthogonal Ensemble (JOE) with contour $\gamma:=[0,1]$.

\subsubsection{Matrix integral solutions (II)}
As was shown in \cite{kakei00,HZL}, the partition function of the symplectic ensemble can also play a role as the tau function of the Pfaffianized system. Therefore, let's consider the case $\beta=4$ with regarding to the equation \eqref{1.1}. The form of the partition function is
\begin{align*}
&\mathbb{Z}_{N}^{(\beta=4)}=\frac{1}{N!}\int_{\gamma^{N}}
\prod_{1\le i<j\le N}(t_i-t_j)^4\prod_{i=1}^{N}\omega^2(t_i)dt_i
\end{align*}
without any constraint on $N$.
By making the use of two-fold Vandermonde determinant
\begin{equation*}
\prod_{1\leq j<k\leq N}(x_j-x_k)^4=\det[x^j_k,(j-1)x_k^j]_{j=1,\dots,2N,k=1,\dots,N}
\end{equation*}
as well as the de Bruijn formula, the partition function above can be written in terms of Pfaffian as
\begin{align*}
\Pf\left((j-i)\int_\gamma t^{i+j-2}\omega(t)dt
\right)_{i,j=1}^{2N}.
\end{align*}
The parameter dependent Pfaffian elements can be chosen as
\begin{align*}
\pf(i,j)=(j-i)\int_\gamma t^{i+j-2}\omega(t;x,z,m,n)dt,
\end{align*}
where the weight function should solve the conditions \eqref{drpf1}-\eqref{drpf3}. A suitable choice of the weight function is
\begin{align*}
\omega(t;x,z,m,n)=t^{n}(1-t)^{-m/2}\exp(\eta_1(x,z,t)),\quad \eta_1(x,z,s)=-2zs^{-1}+(-4s^{-2}+4s^{-1})x+\eta_0(s)
\end{align*}
where $\eta_0(s)$ is taken to ensure the convergence. Obviously, the choice of the weight (dispersion relation) in the $\beta=4$ case is the same as the choice in $\beta=1$ case. Moreover, we can check that
\begin{align*}
& f^n_m=\pf(1,2,\dots,2N)^n_m=\mathbb{Z}_N^{(\beta=4)}(x,z,m,n),\\
& g^n_m=\pf(2,\dots,2N-1)^n_m=\mathbb{Z}_{N-1}^{(\beta=4)}(x,z,m,n+1),\\
& h^n_m=\pf(0,1,\dots,2N,2N+1)^n_m=\mathbb{Z}_{N+1}^{(\beta=4)}(x,z,m,n-1).
\end{align*}
It's also interesting to point out that the tau function $f_m^n$ in this case is related to the partition function of the Jacobi Symplectic Ensemble (JSE)
\begin{align*}
\mathbb{Z}_N^{(\beta=4)}(x,z,m,n)=\frac{1}{N!}\int_{\gamma^N}\prod_{1\leq i<j\leq N}(t_i-t_j)^4\prod_{i=1}^N t_i^{2n} (1-t_i)^{-m}\exp(2\eta_1(x,z,t_i))dt_i,
\end{align*}
with contour $\gamma:=[0,1]$.

\subsection{matrix integral solutions of the Pfaffianized Leznov lattice}
In this subsection, we consider the matrix integral representation for solutions of the Pfaffianized Leznov lattice. The Pfaffianized  Leznov lattice
\begin{align}
& (D_yD_z-2(e^{D_n}-1))f_n\cdot f_n=-2g_n\hat{g}_n,\\
& (D_xD_y-2D_ze^{D_n})f_n\cdot f_n=2D_z g_n\cdot \hat{g}_n,\\
& D_ye^{D_n/2}g_n\cdot f_n=-D_ze^{D_n}g_n\cdot f_n,\\
&  D_ye^{D_n/2}f_n\cdot \hat{g}_n=-D_ze^{D_n}f_n\cdot \hat{g}_n.
\end{align}
was derived in \cite{zhao}. The solution was presented in Pfaffian form
\begin{align*}
& f_n=\pf(1,2,\cdots,N),\\
& g_n=\pf(0,1,\cdots,N+1),\\
& \hat{g}_n=\pf(2,3,\cdots,N-1),
\end{align*}
where the Pfaffian entries satisfy the relation
\begin{align}
& \frac{\partial}{\partial y}\pf(i,j)=\pf(i+1,j)+\pf(i,j+1),\quad \frac{\partial}{\partial z}\pf(i,j)=-\pf(i-1,j)-\pf(i,j-1),\\
& \frac{\partial}{\partial x}\pf(i,j)=-\pf(i-2,j)-\pf(i,j-2),\quad (i,j)_{n+1}=\pf(i+1,j)+\pf(i,j+1).
\end{align}

\subsubsection{ Matrix integral solutions (I)}
Similar to the semi-discrete case,  we first consider the matrix integral related to the orthogonal ensemble with even size $N$, which is of the form
\begin{align*}
	\mathbb{Z}_N^{(\beta=1)}(x,y,z,n)&=
	\frac{1}{N!}\int_0^{\infty}\dots \int_0^{\infty} \prod_{i=1}^N t_i^n \prod_{1\le i<j\le N}|t_i-t_j|\exp[\sum_{i=1}^N \eta_1(x,y,z,t_i)]dt_1 \dots dt_N \notag \\
	&=\Pf\Big[\int_0^{\infty} \int_s^{\infty} (s^{n+i-1}t^{n+j-1}-s^{n+j-1}t^{n+i-1})\exp[\eta_1(x,y,z,s)+\eta_1(x,y,z,t)]dsdt\Big]_{i,j=1,\dots,N}
	\end{align*}
where $\eta_1 (x,y,z,t)=-zt^{-1}-xt^{-2}+ty+ \eta_0(t)$. By using the de Bruijn formula backwards, we find that
\begin{align*}
& f^n_m=\pf(1,2,\dots,N)^n_m=\mathbb{Z}_N^{(\beta=1)}(x,z,t,n),\\
& g^n_m=\pf(0,1,2,\dots,N+1)^n_m=\mathbb{Z}_{N+2}^{(\beta=1)}(x,z,t,n-1),\\
& h^n_m=\pf(2,\dots,N,N-1)^n_m=\mathbb{Z}_{N-2}^{(\beta=1)}(x,z,t,n+1).
\end{align*}

\subsubsection{ Matrix integral solutions (II)}
By using the two-fold Vandermonde determinant and the de Bruijn formula, we can check that
\begin{align*}
	\mathbb{Z}_N^{(\beta=4)}(x,y,z,n)&=
	\frac{1}{N!}\int_0^{\infty}\dots \int_0^{\infty} \prod_{i=1}^N t_i^n \prod_{1\le i<j\le N}|t_i-t_j|^4
	\exp[\sum_{i=1}^N 2\eta_1(x,y,z,t_i)]dt_1 \dots dt_N \notag\\
	&=\Pf\Big[\int^{\infty}_0(i-j)t^{n+i+j-2}\exp[\eta_1(x,y,z,t)]dt\Big]_{i,j=1}^{2N}
	\end{align*}
gives Pfaffian solutions
\begin{align*}
 & f_n=\pf(1,2,\cdots,2N)=\mathbb{Z}_N^{\beta=4}(x,y,z,n), \\
 & g_n=\pf(0,1,2,\cdots,2N+1)=\mathbb{Z}_{N+1}^{\beta=4}(x,y,z,n-1),\\
 & h_n=\pf(2,3,\cdots,2N-1)=\mathbb{Z}_{N-1}^{\beta=4}(x,y,z,n+1).
\end{align*}
Here we take the weight function as
\begin{align*}
\omega(x,y,z,t)=t^n\exp(\eta_1 (x,y,z,t)),\quad \eta_1(x,y,z,t)=-zt^{-1}-xt^{-2}+ty+ \eta_0(t),
\end{align*}
where $\eta_0(t)$ is taken to ensure the convergence.

\section{Conclusion and discussion}

In this paper, we demonstrated that the Leznov equation, as a generalized $2+1$ dimensional integrable system, admits a matrix integral solution. As is known, there are not many results about the matrix integral solution to the $2+1$ dimensional lattice equation as well as the discrete versions of the lattice equation, therefore, we give more examples to connect the $2+1$ dimensional lattice equation with random matrix theory. In particular, the Jacobi $\beta$-ensemble seems to be fundamental when we discuss about the discretization in the $x$-direction. Moreover, we consider the tau functions of discrete Leznov equation in $x/y$-direction respectively and a full-discrete lattice. The matrix integral solution of full discrete lattice is also rare to see but we have given some hints to see this kind of structure. Orthogonal polynomials with three type deformed weights, the Jacobi type, the Laguerre type and the weights deformed by the interval indicator function, are investigated in \cite{Yang}, especially the relation with Heun equations when degree $n$ is large. It is natural to find the related orthogonal polynomials associated with the deformed weights in the obtained matrix integrals here.

\section*{Acknowledgements}
Authors would like to thank Dr. Shi-Hao Li for helpful discussion and suggestion.
The work  is supported by National Natural Science Foundation of
China (Grant no.11871336).

\vskip .3cm
\appendix

\end{document}